\documentclass[3p,times,procedia]{elsarticle}
\usepackage{nupha_ecrc}

\volume{00}

\firstpage{1}

\journalname{Nuclear Physics A}

\runauth{Bo-Wen Xiao}

\jid{nupha}

\jnltitlelogo{Nuclear Physics A}

\usepackage{amssymb}
\usepackage[figuresright]{rotating}

\begin{document}

\begin{frontmatter}



%% Instructions from Editor: Please use the following \dochead only in the preprint version (e-print arXiv etc.); 
%% use empty \dochead{} when submitting to Nuclear Physics A!
\dochead{XXVIth International Conference on Ultrarelativistic Nucleus-Nucleus Collisions\\ (Quark Matter 2017)}
%% e.g. \dochead{17th International Conference on Dynamical Processes in Excited States of Solids}

\title{Low-$x$ Physics in $pA$ Collisions and at the EIC}


\author{Bo-Wen Xiao}

\address{Key Laboratory of Quark and Lepton Physics (MOE) and Institute
of Particle Physics, Central China Normal University, Wuhan 430079, China}

\begin{abstract}
In this proceeding, I review what we have learned, and what we will be able to learn, about low-$x$ physics from proton-nucleus ($pA$) collisions at RHIC and the LHC, and how this complements future Electron-Ion Collider (EIC) measurements. EIC not only will provide us essential and complementary information about the gluon saturation to the knowledge that we have learned from $pA$ collisions at RHIC and the LHC, it will also allow us to visualize the internal structure of protons and nuclei in a multi-dimensional fashion 
with unprecedented precision. 
\end{abstract}


\end{frontmatter}

% \linenumbers

%\section{}
%\label{}

\section{Introduction}
\label{intro}

In small-$x$ physics, one of the major research goal is to investigate the properties of QCD matter at extremely high gluon density in high energy collisions. Saturation physics formalism\cite{Gribov, Mueller}, which is also known as Color Glass Condensate (CGC)\cite{MVmodel}, can be viewed as the high energy effective theory which describes the QCD dynamics when gluon density becomes so high that non-linear effects, namely, the multiple scattering and non-linear small-$x$ evolution, play an important role. First, we know from HERA (Hadron Elektron Ring Anlage) that gluon density rises rapidity at low-$x$. The intuitive theoretical explanation of such rapid rise is that the probability of the bremsstrahlung radiation of a soft gluon is enhanced logarithmically, which can be cast into the gluon evolution by resumming large small-$x$ logarithms up to arbitrary order. It is then conceivable that gluons can start to overlap and recombine when too many gluons are squeezed in a confined hadron. At the end of the day, this can lead to non-linear QCD dynamics and therefore gluon saturation. The non-linear small-$x$ evolution equations, such as the Balitsky-Kovchegov equation\cite{BK}, or the Jalilian-Marian, Iancu, McLerran, Weigert, Leonidov, and Kovner (JIMWLK) equation\cite{JIMWLK}, resums small-$x$ logarithms and takes the saturation effects into account in the meantime. In saturation physics, we always introduce a characteristic scale, namely, $Q_s(x)$ to separate the saturated dense regime from the dilute regime in the $Q^2-x$ plane. One can also interpret $Q_s(x)$ as the typical transverse momentum of gluon at given $x$, with $x$ being the longitudinal momentum fraction w.r.t. its originating hadron. 

Recently, accompanied by tremendous experimental efforts, there have been quite a lot of progress in the small-$x$ field in both theory and phenomenology, which in turn allows us to suggest and propose more interesting measurements at the future Electron-Ion Collider (EIC)\cite{Boer:2011fh, Accardi:2012qut} to search for the decisive evidence for gluon saturation. In particular, we solved the longtime puzzle of why unintegrated gluon distributions are not unique in small-$x$ physics. The ongoing $pA$ program detailed in the RHIC Cold QCD plan\cite{Aschenauer:2016our}, which can serve as a portal to the EIC, also plays an important and complementary role in the study of gluon saturation. Furthermore, a tight connection between small-$x$ physics formalism and Transverse Momentum Dependent (TMD) factorization has been built up. In addition, a lot of efforts have been devoted to the first complete next-to-leading order (NLO) correction to the forward rapidity single inclusive spectra in proton-nucleus collisions and a sophisticated numerical program called Saturation physics at One Loop Order, or SOLO has been developed, in order to numerically study hadron productions up to NLO accuracy. Last but not least, theorists have been working closely with experimentalists on the research and development of the future EIC and have provided rich quantitative and novel predictions of experimental signature for gluon saturation effects.

\begin{figure}[t]
\begin{center}
\includegraphics[width=8cm]{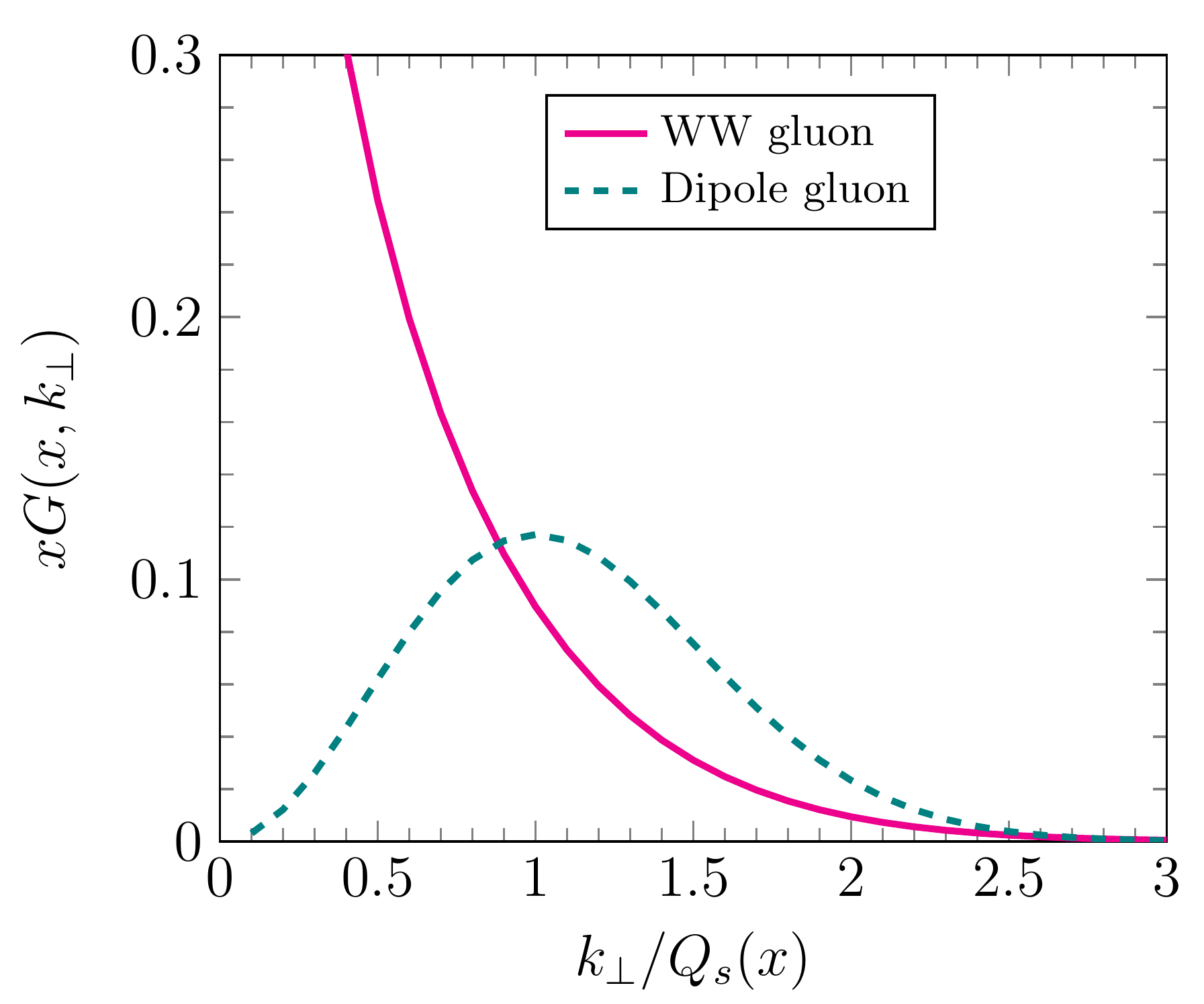}
\end{center}
\caption[*]{The transverse momentum distributions of the Weizs\"{a}cker-Williams and dipole gluon distributions. These two distributions exhibit distinct $k_\perp$ behaviour in the saturation regime, although they have the same normalization after integrating over $k_\perp$ and the same perturbative tale.}
\label{2g}
\end{figure}

\subsection{A tale of two gluon distributions}
For more than a decade, there had been a puzzle of two gluon distributions in small-$x$ physics. On one hand, the dipole unintegrated gluon distribution (UGD), which is defined as the Fourier transform of dipole-target cross section, often appears in the small-$x$ calculation for various inclusive and semi-inclusive processes. In the early days of low-$x$ physics research, it was widely believed that the gluon distribution is unique and universal. What was normally used in various calculations is the dipole-gluon distribution. On the other hand, another different UGD, which is known as the Weizs\"{a}cker-Williams gluon distribution, has also been uncovered in the year around 2000. Regarded as the genuine number density of gluon inside a target hadron, the WW gluon distribution is computed by using the well-known WW approximation and treating the large-$x$ quarks as the source for low-$x$ gluon quanta. (Such method was used to derive the quasi-real photon distribution in Jackson's `bible' on the classical electrodynamics. The WW gluon distribution can be understood as the non-Abelian and non-linear extension of the quasi-photon distribution defined in the classical electrodynamics.\cite{Kovchegov:1996ty}) In the McLerran-Venugopalan model\cite{MVmodel} for a large nucleus, these two UGDs are found to have dramatically different behaviour as function of $k_\perp$, 
\begin{eqnarray}
xG_{\textrm{DP}}(x, k_\perp)&=&\frac{S_{\perp } N_c}{2\pi ^{2}\alpha _{s}} k_\perp^2 \int \frac{d^{2}r_{\perp }}{(2\pi )^{2}}
e^{-ik_{\perp }\cdot r_{\perp }}  e^{-\frac{r_{\perp}^{2}Q_{s}^{2}}{4}}, \\
xG_{\textrm{WW}}(x, k_\perp)&=&\frac{S_{\perp }}{\pi ^{2}\alpha _{s}}\frac{%
N_{c}^{2}-1}{N_{c}} \int \frac{d^{2}r_{\perp }}{(2\pi )^{2}}\frac{%
e^{-ik_{\perp }\cdot r_{\perp }}}{r_{\perp }^{2}}\left[ 1-e^{-\frac{r_{\perp
}^{2}Q_{sg}^{2}}{4}}\right],
\end{eqnarray}
especially in the small $k_\perp$ region, where saturation effects are significant. These $k_\perp$ distributions of these two gluon distributions are illustrated in Fig.~\ref{2g}. Here $Q_{sg}^{2}=\frac{2N_c^2}{N_c^2-1}Q_{s}^{2}$ is the saturation momentum square in the adjoint representation. The above findings had prompted several puzzling yet intriguing questions:
\begin{enumerate}
\item Are UGDs universal? Which gluon distributions are we measuring in a given process? 
\item Why are there exactly two gluon distributions? Are there more UGDs?
\item Is the conventional gluon density, namely the Weizs\"{a}cker-Williams gluon distribution, measurable?
\end{enumerate}
Adapted from the title of the famous classical novel A Tale of Two Cities by Charles Dickens, this puzzle was later dubbed ``A Tale of Two Gluon Distributions" by the authors of Ref.~\cite{tale}. 

This longtime puzzle was solved recently\cite{factorization}. The solution deeply roots in the foundation of QCD that it is a non-Abelian gauge theory, which also requires that gluon distributions must be gauge invariant. Starting from the operator definitions of UGDs, which are the same as the definition of gluon TMDs\cite{Bomhof:2006dp}, we can find that there are only two topologically different gauge invariant and fundamental definitions depending on the choice of future/past gauge links, and these two definitions correspond to the Weizs\"{a}cker-Williams and dipole UGDs, respectively, in the small-$x$ limit.  
Although we confirmed that UGDs are no longer strictly universal, we can show that these two UGDs can be viewed as two fundamental building blocks and we further demonstrate that all of other more complicated UGDs can be constructed from these two universal UGDs in the large $N_c$ limit. The process dependence of UGDs are related to different choices of gauge links. The future and past gauge links correspond to final and initial state interactions, respectively. For a given process, UGDs involved are uniquely determined according to the gauge links associated with the process. As a result, we can build up an effective factorization in small-$x$ physics based on the above generalized universality of UGDs. 

Furthermore, after choosing the light-cone gauge together with proper boundary condition, one can find that the gauge links in the definition of WW gluon distribution completely disappears, which indicates that WW gluon distribution can be interpreted as the genuine gluon density. In contrast, dipole gluon distribution has no such interpretation since it always has gauge link dependence remaining in its definition. In the small-$x$ dipole formalism, which computes all the scattering amplitudes in coordinate space, we find that the WW gluon distribution corresponds to the so-called quadrupole scattering amplitude. The small-$x$ evolution of dipole gluon distribution is governed by the Balitsky-Kovchegov equation or equivalently by the JIMWLK evolution of the dipole scattering amplitude, while the evolution of WW distribution is given by the JIMWLK evolution of the quadrupole amplitude\cite{Dominguez:2011gc}. Close connections\cite{Sudakov, Balitsky:2015qba, Zhou:2016tfe} between the small-$x$ factorization and the TMD factorization have also been established recently, when these two frameworks are both applicable. This gives us a more complete framework in small-$x$ physics to incorporate the Sudakov resummation, namely the parton shower effect, and thus make more reliable quantitative predictions for hard processes, such as forward dijet productions. 

In addition, we find that back-to-back dijet production production processes can help us measure different gluon distributions and tell their difference as summarized in the following table, where $\surd$ and $\times$ indicate that the corresponding gluon distribution appear and do not appear in certain processes, respectively. 
\begin{center}
%\begin{table}[tbp]
\begin{tabular}{|c|c|c|c|c|c|}
 \hline
 &\textrm{Inclusive} & Single inclusive & DIS dijet & $\gamma$ +jet & dijet in pA \\
 \hline $xG_{\textrm{WW}}$& $\times $ & $\times$ & $\surd$& $\times $&
$\surd$\\
\hline $xG_{\textrm{DP}}$ &$\surd$ &$\surd$
       &$\times$ & $\surd$& $\surd$ \\
\hline
 \end{tabular}
% \end{table}
 \end{center}
For example, we can directly measure the dipole gluon distribution in $\gamma$ +jet productions in $pA$ collisions and probe it in single inclusive hadron productions in $pA$ collisions as well as in inclusive deep inelastic scattering (DIS) measurement. Interestingly, dijet productions in DIS can provide us the very first and direct measurement of the WW gluon distribution. This process is then immediately identified as one of the golden measurements in the planned EIC. In forward dijet production processes in $pA$ collisions, due to complicated structure of initial and final state interactions, both gluon distributions enter the cross-section and they start to convolute together to generate more complicated forms of gluon distributions. From the perspective of distinguishing and measuring these two fundamental gluon distributions, measurements in $pA$ collisions and at the future EIC are tightly connected with complimentary physics missions.

\section{What we have learnt in DIS and $pA$ collisions}
\label{current}
In the following, let us discuss the small-$x$ physics that we have learnt in DIS at HERA and in $pA$ collisions at RHIC and the LHC. 
\subsection{Geometrical scaling}
At HERA, we learnt that gluons dominate the parton distributions at low-$x$, and the structure function $F_2(x, Q^2)$ exhibits the so-called geometrical scaling behaviour\cite{Stasto:2000er}, which is considered as the one of the most important hint of gluon saturation. Normally, DIS structure functions is plotted as function of $x$ and $Q^2$, independently. The geometrical scaling phenomenon shows that all the low-$x$ ($x<10^{-2}$) and low-$Q^2$ ($Q^2 < 450 \textrm{GeV}^2$) data points measured at HERA remarkably fall on a single curve of variable $\tau \equiv \frac{Q^2}{Q_s^2}$ (or they can be described by a single function of $\tau$) with $Q_s^2=\left(\frac{x_0}{x}\right)^\lambda \textrm{GeV}^2$, $x_0=3.04\times 10^{-3}$ and $\lambda=0.288$. This phenomenon can be naturally explained as a result of the traveling wave properties of the solution\cite{Munier:2003vc} to the non-linear BK equation. The non-linear mathematical structure of the BK equation, which governs the small-$x$ evolution of dipole scattering amplitudes, dictates the traveling wave solution at asymptotic high energy (for a pedagogical review, see Ref.~\cite{Munier:2014bba}). The geometrical scaling can be simply derived from the traveling wave solution, since the latter only depends on a single variable of $\ln Q^2 -\ln Q_s^2(x)$. Such scaling, which holds for the dipole scattering amplitude, can be carried over to the DIS structure function $F_2(x, Q^2)$ at low-$x$. 

\subsection{Forward single inclusive hadron productions in $pA$ collisions} 
In the process of single inclusive hadron productions in forward $pA$ collisions, the forward hadron is produced from the large-$x$ parton in the proton wave-function after multiple scattering with the dense small-$x$ gluons inside the target nucleus wave-function. During the multiple scattering, the forward moving parton receives typical transverse momentum of order of $Q_s$ from the target. By measuring the $p_\perp$ spectra of forward rapidity hadrons, we can indirectly study the saturation momentum in the low-$p_\perp$ region\cite{Dumitru:2002qt}. By going to forward rapidity region, we can take the advantage of the kinematics which limits the parton from the proton projectile in the large-$x$ region, while it also ensures that the gluons from the target nucleus are deeply in the low-$x$ region. The leading order contribution of this process in the small-$x$ formalism together with some NLO effects, such as the running coupling effect\cite{Kovchegov:2006vj, Balitsky:2008zza} and $\alpha_s$ corrections, have been studied quite extensively\cite{Dumitru:2005gt, Albacete:2010bs, Levin:2010dw, Fujii:2011fh, Albacete:2012xq, Lappi:2013zma}.

At NLO, extra gluon radiations can also affect the transverse momentum of the produced hadron\cite{Altinoluk:2011qy}. This implies that the NLO correction can be important since it opens up new channels and brings additional source to the transverse momentum of the produced hadron.  Recently, we have taken an important first step towards the NLO phenomenology in the saturation formalism. In Refs.~\cite{NLO, NLO-numerical}, we have not only computed the first complete NLO correction to the forward single inclusive hadron spectrum in $pA$ collisions, but also developed a rather sophisticated numerical package (SOLO) to evaluate the NLO hadron spectra and found excellent agreement with RHIC\cite{Arsene:2004ux, Adams:2006uz} and the LHC\cite{Milov:2014ppp} data in the low $p_\perp$ region, where saturation effects play an essential role while the conventional collinear formalism becomes inadequate or even breaks down\footnote{The collinear framework always assumes that incoming partons carry no transverse momentum, and it becomes unreliable in the region where hadron $p_\perp$ is small.}. This therefore provides a precise and reliable test of saturation physics beyond the leading logarithmic approximation. 

Due to large and negative NLO corrections, SOLO breaks down\cite{NLO-numerical} in the large $p_\perp$ region where $p_\perp > (2 \sim 3) Q_s(x)$, which is in principle already outside the region of validity of saturation formalism. When the transverse momentum of the produced hadron $p_\perp$ gets large, hard collisions becomes the dominant source of hadron $p_\perp$, while the saturation effects are expected to be negligible. As a result, the collinear formalism can naturally describe the data in the large $p_\perp$ region as expected\cite{Stasto:2014sea}. Nevertheless, there have been also a lot of efforts\cite{Beuf:2014uia, Altinoluk:2014eka, Kang:2014lha, Ducloue:2016shw, Stasto:2016wrf, Iancu:2016vyg, Ducloue:2017mpb} trying to extend the NLO calculation in the saturation formalism to larger $p_\perp$ region and build a more complete framework.

\subsection{Dihadron productions in $pA$ collisions}

Forward rapidity back-to-back dihadron (dijet) productions in $pA$ collisions have been considered as one of most interesting probes which are directly sensitive to the gluon saturation effects in the nucleus target. One of the main sources of this back-to-back correlation in the forward rapidity region is due to the transverse momentum (of order $Q_s$) carried by the low-$x$ gluons in the target nucleus. As compared to the same observable in $pp$ collisions (or peripheral $dAu$ collisions), one expects that there should be more suppression in the back-to-back region since the saturation momentum $Q_s(A)$ for a target nucleus is roughly $A^{1/3}$ time of that for a proton. The quantitative feature of this de-correlation was first computed and predicted by Ref.~\cite{Marquet:2007vb}. There have been some interesting experimental evidences \cite{Braidot:2010ig, Adare:2011sc, Li:2012bn} on the suppression of the back-to-back dihadron correlations at forward rapidity in $dAu$ collisions at RHIC. 

As shown in Ref.~\cite{factorization}, both the WW and dipole gluon distributions are involved in the back-to-back dihadron (dijet) productions in $pA$ collisions, due to complicated structure of the initial and final state effects. The numerical studies\cite{Albacete:2010pg, Stasto:2011ru, Lappi:2012nh} of the forward dihadron correlation with slightly different modelling of saturation effects find good agreement with the experimental data\cite{Braidot:2010ig, Adare:2011sc}. The physical picture of this process at partonic level is that a large-$x$ parton coming from the proton projectile splits into two partons before or after interacting with the dense nuclear medium and eventually produces a pair of back-to-back jets at forward rapidity $y$. Although the transverse momentum of each jet $P_{i\perp}$ is large, the transverse momentum imbalance $q_\perp=|P_{1\perp}+P_{2\perp}|$ of these two jets remains relatively small, and it mainly comes from the small-$x$ gluon with transverse momentum $k_\perp$ originated from the target nucleus. The transverse momentum of small-$x$ gluon distribution $k_\perp$ typically is of the order of the saturation momentum $Q_s$, which is a scale characterising the strength of the saturation effect. When the saturation effect is small, $Q_s$ is small. Therefore, we expect a strong peak in dijet (dihadron) back-to-back correlations. On the other hand, when the saturation effect becomes strong, we anticipate that the dijet momentum imbalance $q_\perp$ gets large, and thus the away side ($\Delta \phi \simeq \pi$) peak gets suppressed. By measuring the dijet back-to-back azimuthal angle $\Delta \phi \simeq \pi$ correlation, one can probe the typical transverse momentum of small-$x$ gluons, therefore extract information about the saturation effects inside target nucleus. Using the small-$x$ improved TMD factorization framework, a significant suppression of the forward di-jet angular correlations in proton-lead versus proton-proton collisions at the LHC due to saturation effects has been predicted in Ref.~\cite{vanHameren:2016ftb}. Besides the above calculations, there are also other explanations, which is not based on small-$x$ saturation formalism. For example, studies in Refs.~\cite{Qiu:2004da, Kang:2011bp} show that the dihadron correlations measured in $dAu$ collisions by both PHENIX and STAR collaborations at RHIC can be described by the cold nuclear matter energy loss effects and coherent power corrections.

So far, the phenomenological studies of the di-hadron correlations in $pA$ collisions are mostly based on the LO results in the small-$x$ formalism. Also, we need to assume that the initial transverse momentum of partons coming from the projectile proton is negligible. To make more precise and reliable predictions for $pAu$ collisions at RHIC, one needs to go beyond LO by including the one-loop contributions, especially the parton shower effect. In Ref~\cite{Sudakov}, we explicitly demonstrated that one can perform the Sudakov resummation (also known as the Collins-Soper-Sterman (CSS) resummation\cite{Collins:1984kg}) in small-$x$ formalism when both small-$x$ and Sudakov type logarithms are important. This study provides us deep insights into the understanding of factorization for high energy hard scatterings in the small-$x$ formalism, and shows that one can incorporate the TMD (CSS) evolution in the small-$x$ formalism for hard processes. We have also suggested that the Sudakov resummation, which is equivalent to the parton shower used in Monte Carlo generators, can also play an important role in back-to-back dijet angular correlations in general. More sophisticated numerical computation which includes both saturation effects and parton shower effects shall be available in the near future and detailed comparisons with the new forward dihadron data measured in $pAu$ collisions at RHIC will be carried out and released soon. This eventually may provide us another piece of strong evidence for the gluon saturation phenomenon.

\section{What we will be able to learn in the future at EIC}
\label{future}

Strongly recommended by the nuclear science advisory committee in the US as the next generation high energy nuclear physics facility, the proposed cutting-edge EIC can lead us to answers to many fundamental questions about the physical role and three-dimensional image of gluons in nucleons and nuclei with unprecedented precision, and also has the potential to discover an interesting form of ultra-dense gluonic matter at the onset of the gluon saturation phenomenon. In particular, EIC will be a fascinating ``stereoscopic camera" with excellent resolution, which allows us to visualise protons and nuclei in a multi-dimensional fashion and provides us the transverse momentum and impact parameter dependence of various quark and gluon distributions. 

\begin{figure}[t]
\begin{center}
\includegraphics[width=14cm]{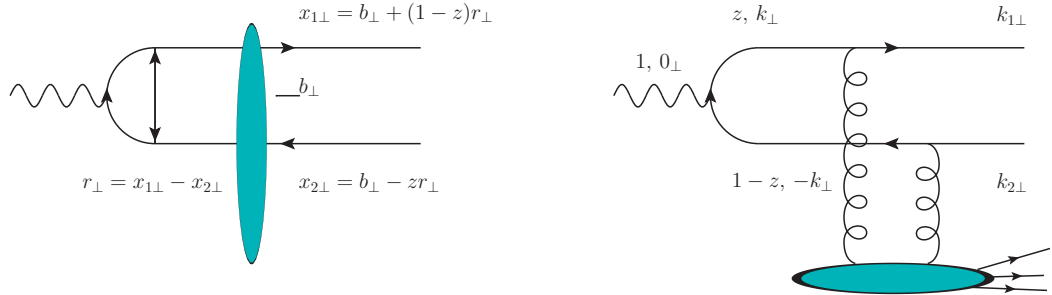}
\end{center}
\caption[*]{Dijet productions in deep inelastic scatterings in coordinate space (left figure) and momentum space (right figure).}
\label{dis2j}
\end{figure}

As one of the most interesting channels, the DIS dijet productions as schematically shown in Fig.~\ref{dis2j} are sensitive to the WW gluon distribution and allow us to directly measure this genuine gluon density for the first time\cite{factorization}. The momentum imbalance of the produced back-to-back dihadron pairs are strongly affected by the transverse momentum $k_\perp$ of the WW gluon distribution. By combining measurements conducted in various processes in $pA$ collisions, we will be able to tell the characteristic difference (as illustrated in Fig.~\ref{2g}) between the WW gluon distribution and the dipole gluon distribution, which is regarded as one of the signature predictions of small-$x$ physics in the saturation regime. Also, by comparing the back-to-back dihadron angular correlation in $ep$ and $eA$ collisions, we have the opportunity to search for the clues of gluon saturation phenomenon with high precision\cite{Zheng:2014vka}. One can further study the azimuthal elliptic anisotropy\cite{Dumitru:2015gaa} of the dijet system in DIS, which can reveal interesting information about the linearly polarized gluon UGD at small-$x$\cite{Metz:2011wb, Dominguez:2011br}. 

Among various kinds of parton distributions, the quantum phase space Wigner distribution\cite{Ji:2003ak, Belitsky:2003nz} is of the greatest importance, since it ingeniously encodes all the quantum information of how partons are distributed inside hadrons. From the Wigner distribution, we can obtain the momentum distributions of quark and gluon together with the information of their impact parameter dependence. One can also define the so-called generalized TMD\cite{Meissner:2009ww}, which is fully written in momentum space, as the Fourier transform of the Wigner distribution. Furthermore, TMDs and generalized parton distributions (GPDs) appear as certain limits of the generalized TMDs or Wigner distributions. 

In the past, it was believed that the parton Wigner distributions probably can not be directly measured in high energy scatterings. Recently, new progress in the study of the gluon Wigner distribution shows that, as a matter of fact, the dipole type Wigner gluon distribution (the dipole type generalized TMD to be more precise) in low-$x$ region is equivalent to the impact parameter dependent dipole gluon distribution, and it can be measured\cite{Wigner} in diffractive dijet processes\cite{Nikolaev:1994cd, Bartels:1996ne, Bartels:1996tc, Diehl:1996st, Altinoluk:2015dpi} at EIC. Further efforts along this direction shall enable us to conduct three-dimensional tomography of proton in the low-$x$ region in the EIC era. The experimental signature of this diffractive process is the large rapidity gap between the produced dijet system and target proton/nucleus, which remains intact with momentum recoil $\Delta_\perp$. The distribution of the large jet momentum and the momentum recoil $\Delta_\perp$ are related to the momentum distribution and impact parameter ($b_\perp \sim 1/\Delta_\perp$ is conjugate to $\Delta_\perp$) distributions of the low-$x$ gluon, respectively.

In addition, other processes, such as diffractive vector meson productions in DIS\cite{Brodsky:1994kf, Kowalski:2003hm, Lappi:2014foa, Mantysaari:2016ykx} and deeply virtual Compton scattering\cite{Ji:1996nm, Radyushkin:1997ki, Kowalski:2006hc, Hatta:2017cte} in the low-$x$ limit, give us the spatial image of gluons inside hadrons, which is also known as the gluon GPDs. 

\section{Conclusion}
\label{conc}

Recently, with the help of much experimental efforts especially in $pA$ collisions at RHIC and the LHC, a lot of progress in low-$x$ physics has been achieved towards better theoretical understanding and more precise phenomenological description of various observables. In particular, we have more precise description of the single inclusive hadron and dihadron productions in forward $pA$ collisions. Complementary measurements in $pA$ collisions at RHIC and the LHC as well as in DIS at the future EIC can lead us to the measurement of both WW and dipole gluon distributions as well as gluon Wigner distributions at low-$x$. Ultimately, the proposed cutting-edge EIC will allow 
mankind to depict the three-dimensional colorful landscape of the internal structure of proton and nucleus at low-$x$.

\textbf{Acknowledgements}:
This material is based upon work supported by the Natural Science Foundation of China (NSFC) under Grant No.~11575070.



%% References


\bibliographystyle{elsarticle-num}



\end{document}